\documentclass[aps,floatfix,superscriptaddress,twocolumn,tightenlines,amsmath,amssymb,nofootinbib,10pt,longbibliography,prl]{revtex4-2}

\usepackage[utf8]{inputenc}
\usepackage[T1]{fontenc}
\usepackage[mathscr]{euscript}
\usepackage{graphicx}
\usepackage[dvipsnames]{xcolor}
\usepackage{amsthm}
\usepackage{psfrag}
\usepackage{ifsym}
\usepackage{dsfont}
\usepackage{enumerate}
\usepackage{amsfonts}
\usepackage{multirow}
\usepackage{mathtools}
\usepackage{amsmath}
\usepackage{hyperref}
\hypersetup{colorlinks=true, linkcolor=red, citecolor=blue}
\usepackage[sort&compress]{natbib}
\usepackage[separate-uncertainty=true]{siunitx}
\usepackage{comment}
\usepackage[normalem]{ulem}
\usepackage[sc,osf]{mathpazo}
\usepackage{bm}
\usepackage{bbm}
\usepackage{xr}

\newtheorem{theorem}{Theorem}

\newcommand{\bra}[1] {\langle #1 |}
\newcommand{\ket}[1] {| #1 \rangle}

\newcommand{\fref}[1]{Fig.~\ref{#1}}
\newcommand{\id}{\mathds{1}}

\def\1{\mathbf{1}}
\def\0{\mathbf{0}}

\def\st{\textrm{subject to }}

\DeclareMathOperator*{\maximize}{Maximize}

\def\P{\mathcal{P}}

\DeclareMathOperator{\Tr}{Tr}

\def\y{\mathbf{y}}

\begin{document}

\title{Quantum communication complexity beyond Bell nonlocality}

\author{Joseph Ho}
\affiliation{Institute of Photonics and Quantum Sciences, School of Engineering and Physical Sciences, Heriot-Watt University, Edinburgh EH14 4AS, United Kingdom}

\author{George Moreno}
\affiliation{International Institute of Physics, Federal University of Rio Grande do Norte, 59070-405 Natal, Brazil}

\author{Samuraí Brito}
\affiliation{International Institute of Physics, Federal University of Rio Grande do Norte, 59070-405 Natal, Brazil}

\author{Francesco Graffitti}
\affiliation{Institute of Photonics and Quantum Sciences, School of Engineering and Physical Sciences, Heriot-Watt University, Edinburgh EH14 4AS, United Kingdom}

\author{Christopher L. Morrison}
\affiliation{Institute of Photonics and Quantum Sciences, School of Engineering and Physical Sciences, Heriot-Watt University, Edinburgh EH14 4AS, United Kingdom}

\author{Ranieri Nery}
\affiliation{International Institute of Physics, Federal University of Rio Grande do Norte, 59070-405 Natal, Brazil}

\author{Alexander Pickston}
\affiliation{Institute of Photonics and Quantum Sciences, School of Engineering and Physical Sciences, Heriot-Watt University, Edinburgh EH14 4AS, United Kingdom}

\author{Massimiliano Proietti}
\affiliation{Institute of Photonics and Quantum Sciences, School of Engineering and Physical Sciences, Heriot-Watt University, Edinburgh EH14 4AS, United Kingdom}

\author{Rafael Rabelo}
\affiliation{Instituto de Fisica ``Gleb Wataghin'', Universidade Estadual de Campinas, 13083-859, Campinas, Brazil}

\author{Alessandro Fedrizzi}
\affiliation{Institute of Photonics and Quantum Sciences, School of Engineering and Physical Sciences, Heriot-Watt University, Edinburgh EH14 4AS, United Kingdom}

\author{Rafael Chaves}
\affiliation{International Institute of Physics, Federal University of Rio Grande do Norte, 59070-405 Natal, Brazil}
\affiliation{School of Science and Technology, Federal University of Rio Grande do Norte, 59078-970 Natal, Brazil}

\begin{abstract}
Efficient distributed computing offers a scalable strategy for solving resource-demanding tasks such as parallel computation and circuit optimisation.
Crucially, the communication overhead introduced by the allotment process should be minimised---a key motivation behind the communication complexity problem (CCP).
Quantum resources are well-suited to this task, offering clear strategies that can outperform classical counterparts.
Furthermore, the connection between quantum CCPs and nonlocality provides an information-theoretic insights into fundamental quantum mechanics.
Here we connect quantum CCPs with a generalised nonlocality framework---beyond the paradigmatic Bell's theorem---by incorporating the underlying causal structure, which governs the distributed task, into a so-called nonlocal hidden variable model.
We prove that a new class of communication complexity tasks can be associated to Bell-like inequalities, whose violation is both necessary and sufficient for a quantum gain.
We experimentally implement a multipartite CCP akin to the guess-your-neighbour-input scenario, and demonstrate a quantum advantage when multipartite Greenberger-Horne-Zeilinger (GHZ) states are shared among three users.
\end{abstract}

\maketitle

Quantum technology enables applications ranging from fundamentally secure cryptography~\cite{gisin2002quantum} to quantum teleportation~\cite{bennett1993teleporting,ren2017ground} and ultimately the quantum internet as enabled by distribution of global-scale entanglement resources~\cite{kimble2008quantum,wehner2018quantum,brito2019statistical}.
Quantum correlations, created in quantum networks, can be harnessed to enhance the efficiency of distributed information processing, i.e., by reducing communication cost; this is exemplified in the use of shared entanglement in communication complexity problems (CCPs)~\cite{Cleve1997,Buhrman2010,PhysRevLett.89.197901}.
This will be of interest in near term quantum computing platforms where many medium-sized nodes are linked to scale up the computational capabilities~\cite{10.1007/978-3-540-45138-9_1,doi:10.1098/rspa.2012.0686}, where CCP will naturally reside.

Distributed computing represents a highly versatile method of solving demanding tasks by taking a global target function and splitting the input among multiple users. The users act on their inputs locally to solve a global problem with some communication allowed between each of the users.
CCP provides the necessary framework in evaluating the ultimate performance of these architectures, notably evaluating the minimum communication overhead needed to achieve the task~\cite{yao1979some,PhysRevA.72.050305,kumar_experimental_2019}.
Recent developments of CCPs have adopted the use of non-classical resources~\cite{Cleve1997}, and an updated definition for evaluating the complexity of a problem.
One might also be interested in obtaining the highest probability of successfully evaluating the target function, with fixed amount of communication~\cite{Brukner2004,Buhrman2010,Buhrman3191,Junge2018fixedCommunication}.
In spite of Holevo's theorem \cite{nielsen2002quantum}---showing that quantum states cannot reduce the cost of transmitting a classical message---if our aim is to compute a function of it, as in the classic CCP setting, quantum resources such as entanglement can demonstrate an improvement.
Remarkably, it has been proven that quantum advantages in CCPs can be mapped to the violation of Bell inequalities \cite{brunner2014bell,Junge2018fixedCommunication}, thus establishing an important link between two key concepts of computer science and quantum theory.
Moreover, it is widely believed that communication complexity should scale with the size of the input data. Interestingly, this non-triviality of a CCP can be seen as an informational principle for why Nature cannot be more nonlocal than what is achievable within quantum theory~\cite{van2013implausible,brassard2006limit}. 

The connection between Bell inequalities and communication complexity has only been proven for the standard notion of Bell nonlocality, which contrasts quantum mechanics with \textit{local} hidden variable (LHV) models.
In this standard scenario, one considers a number of separate parties who share a common source of correlations but cannot communicate---an arrangement that is severely limiting, particularly in the context of CCPs.
The study of Bell nonlocality has however produced much more general and stronger notions of nonlocality~\cite{Svet87,ExperimentalSvetlichny,Chaves2017causalhierarchyof,collins2002bell,bancal2009quantifying,jones2005extent,bancal2011detecting,brask2017bell,chaves2018quantum,poderini_experimental_2020}; these include scenarios that allow subsets of parties involved in a Bell experiment to communicate, while the classical description now considers \textit{non}local hidden variable (NLHV) models.
So far the connection between CCPs and this generalised Bell nonlocality has not been investigated. That is precisely our aim. 

\begin{figure*}
    \includegraphics*[width=\textwidth]{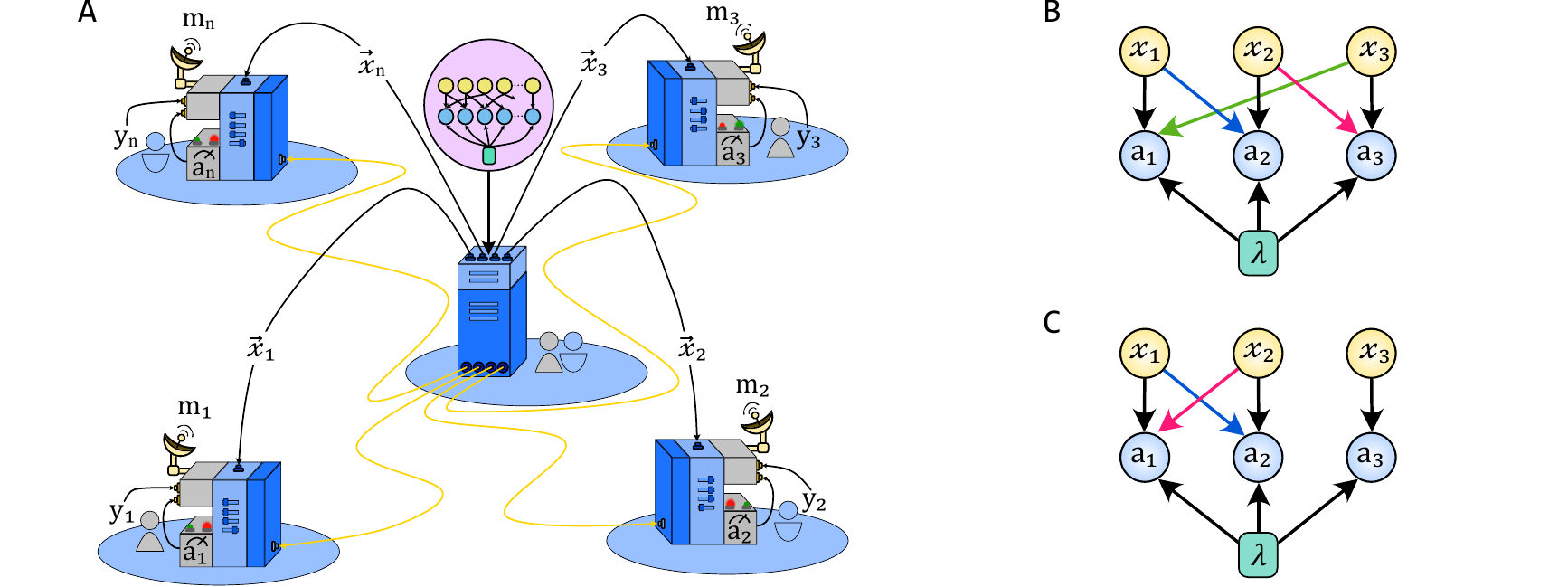}
    \caption[]{\label{fig:CCP}
    \textbf{Distributed computation in a CCP scenario.} 
    \textbf{(A)} A central agent allocates resources among a network of $n$ users who independently produce outcomes to solve a collective task.
    Each user $i$ receives $y_i$ and a subset $\vec{x}_i$ of the variables $\left\{x_1, \cdots, x_n\right\}$ according to some underlying causal structure indicated by the shaded pink region.
    Each user processes their input data along with shared correlations and broadcast a one bit message $m_i$ to all others before they compute a function $f(x_1,\dots,x_n,y_1,\dots,y_n)$.
    When allowing for quantum resources, shared quantum correlations that violate a Bell-like inequality bounding the corresponding NLHV model provide an advantage in the probability of success of the task over its classical counterpart.
    \textbf{(B)} Causal structure of the three-party GYNI scenario \cite{almeida2010guess,Chaves2017causalhierarchyof} experimentally investigated. The input of each party $x_i$ is communicated to its neighbour to the right (or alternatively, to the left). The variable $\lambda$ stands for pre-shared (classical) correlations shared among the parties and used to produces respective outcomes, $a_i$.
    \textbf{(C)} Causal structure of Svetlichny's scenario \cite{Svet87} as discussed in Supplementary Material.
    }
\end{figure*}

Here we show that NLHV models define a new and more general class of CCPs.
As opposed to the standard scenario, where each party has an exclusive part of the input data required to compute the desired function, in the generalised case the input data can be distributed in arbitrary manners.
Every NLHV model defines a specific pattern in which the input data is distributed among the parties, see \fref{fig:CCP}~(A).
Moreover, every full correlator Bell inequality bounding the classical correlations in such NLHV models not only defines a target function in the CCP, but also the corresponding probability of success in trying to compute it with a restricted amount of communication between the parties.
Thus, the violation of these Bell inequalities are a necessary and sufficient condition for a quantum advantage in generalised CCPs.
This establishes the first general connection between this new form of nonclassicality emerging from NLHV models and a relevant quantum information task.
We experimentally investigate a three-party CCP task with a causal structure inspired by the \textit{guess-your-neighbour-input} (GYNI) non-local game, and demonstrate an increased winning probability when using quantum resources which violate the associated Bell-like inequality.
This exemplifies the need to consider the underlying causal structure when defining the Bell-like inequalities for generalised CCPs.

\section{Results}

\textbf{Bell scenarios with communication and nonlocal hidden variable models.} In a standard Bell scenario, each of $n$ distant parties receive an input $x_i$ and output $a_i$ (with $i=1,\dots,n$). A local hidden variable description implies that the observed probability distribution $p(a_1,\dots,a_n \vert x_1,\dots,x_n)=p(\vec{a} \vert \vec{x})$ can be decomposed as
\begin{equation}
\label{eq:lhv}
p(\vec{a} \vert \vec{x})= \sum_{\lambda}p(a_1 \vert x_1,\lambda)\dots p(a_n \vert x_n,\lambda)p(\lambda),
\end{equation}
where $\lambda$ is a classical random variable accounting for all correlations observed between the measurement outcomes of the distant parties.
In turn, in a quantum description, Born's rule implies that
\begin{equation}
\label{eq:quant1}
p(\vec{a} \vert \vec{x})= Tr\left[ \left(M_{a_1}^{x_1} \otimes \dots \otimes M_{a_n}^{x_n} \right) \rho\right],
\end{equation}
where $M_{a_i}^{x_i}$ are measurement operators and $\rho$ describes the quantum state shared between the parties.
As shown by Bell \cite{bell1964einstein}, there are quantum correlations \eqref{eq:quant1} that cannot be written as \eqref{eq:lhv}.
This is the phenomenon known as Bell nonlocality and is witnessed by the violation of Bell inequalities \cite{bell1964einstein,brunner2014bell}, where linear constraints on the probabilities should be respected by any distribution of the form \eqref{eq:lhv}.

In spite of its importance, the usual Bell scenario is rather restrictive, in that no communication can take place between the parties.
Alternatively, one can think of an external agent that generates a sequence of values of $n$ random variables $\left\{x_i\right\}$ and sends (possibly overlapping) subsets of this sequence to each of the $n$ parties involved in the Bell test.
In this more general scenario the $i$-th party can receive a total of $l_i$ inputs that we label as $x_{i,j} \in \left\{x_i\right\}$ with $j=1,\dots,l_i$  and $x_{i,1}=x_i$.
Let the set of inputs of party $i$ be organised in a vector $\vec{x}_{i} = \{x_{i,j}|j=1,\dots,l_{i}\}$.
A classical description is then given by a NLHV model
\begin{eqnarray}
\label{eq:nlhv}
& & p(\vec{a} \vert \vec{x})=  \sum_{\lambda}p(a_1 \vert {\vec{x}_{1}}
,\lambda)\dots p(a_n \vert {\vec{x}_{n}}
,\lambda)p(\lambda),
\end{eqnarray}
that can be graphically represented by a directed acyclic graph where each measurement outcome $a_i$ has a set of parents $x_{i,j}$.
See \fref{fig:CCP}~(B) and \fref{fig:CCP}~(C) for examples.

Analogously to Eq.~\eqref{eq:quant1}, the set of quantum correlations in this extended Bell scenario is described as
\begin{equation}
\label{eq:quant2}
p(\vec{a} \vert \vec{x})= Tr\left[ \left(M_{a_1}^{\vec{x}_{1}} \otimes \dots \otimes M_{a_n}^{\vec{x}_{n}} \right) \rho\right],
\end{equation}
that is, the measurement settings of each party may now depend on subsets of $\left\{x_i \right\}$, denoted as $\vec{x}_{i}$ for the subset held by party $i$, and for different parties these might have an overlap.
From a broad perspective, we are imposing a given causal structure to the experiment, one in which parts of the input of a given party can also be known by other distant parties.
Similarly to the usual Bell's theorem, we will be interested in whether: i) there are quantum correlations, Eq.~\eqref{eq:quant2}, that do not have a classical description as in Eq.~\eqref{eq:nlhv}; and ii) this nonclassicality can be harnessed in the processing of information, in particular in CCPs.

The answer to the first question will inherently depend on the specific causal structure under analysis but positive examples are known \cite{Svet87,Chaves2017causalhierarchyof,collins2002bell,bancal2009quantifying,jones2005extent,bancal2011detecting,brask2017bell,chaves2018quantum} and will be explored in more detail below.
Preceding this, a general answer to the second question is the central theoretical result of this paper.

\medskip
\textbf{Communication Complexity and Bell inequalities.} Without loss of generality, in a usual CCP involving $n$ participants \cite{Brukner2004}, each party $i$ receives two bits---$x_i$ and $y_i$---and can broadcast to all other parties just a one bit message.
As an example of such a CCP, one can consider that the parties want to schedule an appointment, their local inputs represent their availability in different time slots and thus the function they want to compute relates to finding a time slot when all of them are available.
However, one can think of more general scenarios where the schedule (or part of it) from one of the participants is known to the others.
Having this in mind in our generalised CCP each party $i$ has access to the random variables $\vec{x}_{i} = \left\{x_{i,j}|j = 1,\dots,l_{i} \right\} \subset \left\{ x_i | i = 1,\dots,n \right\}$ and $y_i$, where $ x_i,y_i \in \{\pm 1\}$ (see \fref{fig:CCP}~A).
The values of the variables $x_i$ are drawn from a joint probability distribution $q(x_1,\dots,x_n)$, while the $y_i$'s are independently drawn from a uniform distribution.
Furthermore, the parties are also allowed to share correlated systems and use their measurement outcomes $a_i\in \{\pm 1\}$ in the execution of the protocol.
Here, we follow closely the conceptual framework of the seminal results in Ref.~\cite{Brukner2004}, the first to provide a general connection between LHV models and CCPs.
As in Ref. \cite{Brukner2004}, the goal is for each party to evaluate a binary function $f(x_1,...,x_n,y_1,...,y_n)=f(\vec{x},\vec{y})=f$ given by
\begin{eqnarray}
\label{eq:f}
f(\vec{x},\vec{y}) = y_1 \dots y_n S[Q(x_1,\dots,x_n)],
\end{eqnarray}
where $Q = Q(x_{1},\dots,x_{n})$ is a function of all inputs, and $S[Q] = \frac{Q}{|Q|}$ is the sign function,  with the restriction that each party $i$ can only broadcast a single bit $m_{i}=m_{i}(x_{i,1},\dots,x_{i,l_i},y_i,a_i)$ to every other party $j$. If party $i$ guesses $G_i(\vec{x},\vec{y})$ for the function $f(\vec{x},\vec{y})$, its probability of success is given by
\begin{eqnarray}
\mathcal{P}_{i}=\frac{1}{2^{n}}\sum_{\substack{x_1,\dots,x_n \\ y_1,\dots,y_n}}q(\vec{x}) P(G_i(\vec{x},\vec{y})=f(\vec{x},\vec{y})),
\end{eqnarray}
where  $P(G_i(\vec{x},\vec{y})=f(\vec{x},\vec{y}))=1$ if $G_i(\vec{x},\vec{y})=f(\vec{x},\vec{y})$ and $0$ otherwise.

Consider now a general Bell inequality of the form
\begin{eqnarray}
\label{inequality1}
B_n=\sum_{x_1,\dots,x_n=-1}^1 Q(x_1,\dots,x_n) E_{x_1,\dots,x_n}\leq B^C_n,
\end{eqnarray}
where $Q(x_{1},\dots,x_{n})$ is the coefficient of $E_{x_1,...,x_n}$, which stands for the full correlation function
\begin{multline}
\label{eq:fullcorr}
E_{x_1,\dots,x_n}  =  P_{x_1,\dots,x_n}\left(\prod_{i=1}^n a_i=1\right) \\   - P_{x_1,\dots,x_n}\left(\prod_{i=1}^n a_i=-1\right),
\end{multline}
and $B^C_n$ is the classical bound associated with a given causal structure described by the NLHV decomposition in Eq.~\eqref{eq:nlhv}.
Then, our main theoretical result is to show that a violation of such a Bell inequality is necessary and sufficient to lead to a quantum advantage in a CCP related to the computation of the function in Eq.~\eqref{eq:f}.
This is stated in the following theorem, the proof of which is elaborated in the Methods and Supplementary Material.

\begin{theorem}
\label{th:theorem}
Given a Bell inequality of the form \eqref{inequality1}, the optimal classical probability of success $\mathcal{P}^C_i$ of party $i$ computing the function
\begin{eqnarray}
f(\vec{x},\vec{y}) = y_1 \dots y_n S[Q(x_1,\dots,x_n)],
\end{eqnarray}
is limited by 
\begin{eqnarray}
\label{eq:theorem1}
\mathcal{P}^C_{i} \leq\frac{1}{2} + \frac{B^C_n}{2\Gamma}, \;\;\; \forall i;
\end{eqnarray}
with
$\Gamma = \sum_{x_1,\dots,x_n = -1}^1|Q(x_1,\dots,x_n)|$. Moreover, using the correlations shared between the parties there is a protocol achieving
\begin{eqnarray}
\label{eq:theorem2}
\mathcal{P}_{i}=\frac{1}{2} + \frac{B_n}{2\Gamma}, \;\;\; \forall i;
\end{eqnarray}
thus showing that a violation of the Bell inequality \eqref{inequality1} is both necessary and sufficient for an advantage in the CCP.
\end{theorem}

\begin{figure*}[ht!]
    \includegraphics*[width=\textwidth]{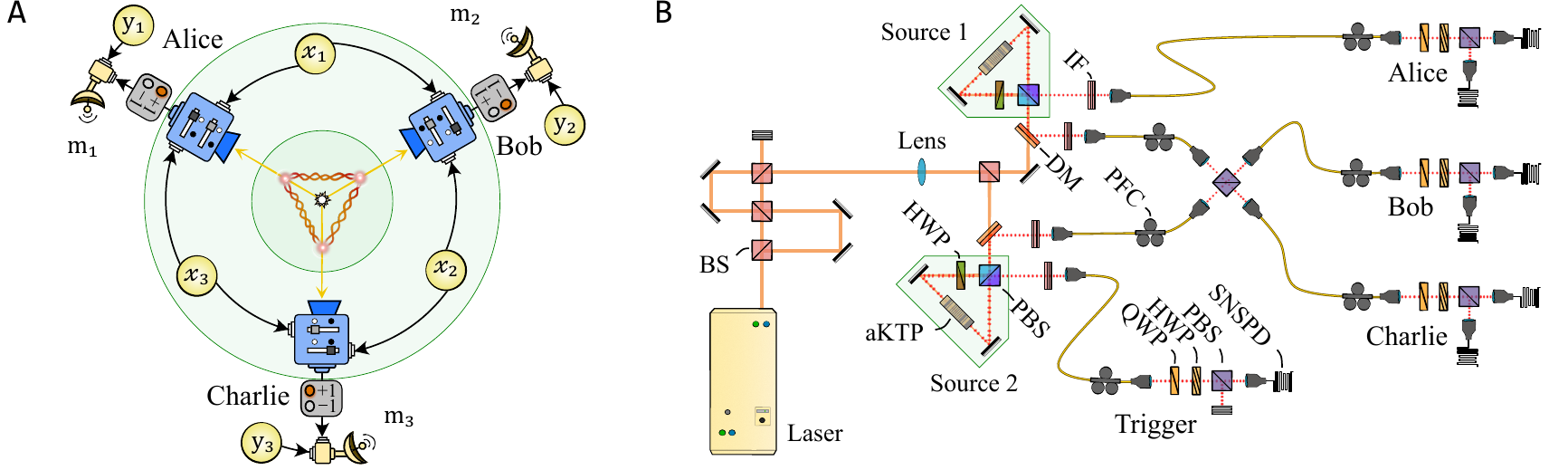}
    \caption{\textbf{Experimental layout.} \textbf{(A)}
    Conceptual layout of the three-user GYNI protocol.
    The tripartite GHZ state is distributed among the three users who locally measure their photon based on input bits \(\{x_i, x_j\}\).
    To compute the target function, each user broadcasts a one-bit message using their measured outcome and a local bit, $y_i$.
    \textbf{(B)} We create the GHZ state using two polarisation-entangled photon-pair sources and a linear-optics fusion gate.
    Each source is implemented with an aperiodically-poled KTP crystal embedded in a Sagnac loop that is optically pumped bidirectionally using a picosecond mode-locked laser, see Methods for details.
    Down-converted photons are separated from the pump laser using dichroic mirrors (DM) and interference filters (IF) then fibre coupled into single mode fibres.
    One photon from each source non-classically interferes on a polarising beamsplitter (PBS) creating the three-photon GHZ state conditioned on measuring the forth photon as a Trigger.
    Each user performs projective measurements on their qubit using a quarter-wave plate (QWP), half-wave plate (HWP), and PBS.
    Single photons are detected using superconducting nanowire single-photon detectors (SNSPD) and time-tagged for coincidence measurements within a \SI{1}{ns} window.}
    \label{fig:expLayout}
\end{figure*}

This result shows that every full correlator Bell inequality that displays a quantum violation is associated to a CCP with quantum advantage, even in scenarios where the parties can communicate.
As detailed in the Methods and Supplementary Material, to achieve the probability of success \eqref{eq:theorem2}, the message $m_i$ communicated from one party to all others should be $m_i=y_ia_i$, that is, the product of its input $y_i$ with the measurement outcome $a_i$.
Interestingly, as we show next, there are inequalities that do not show quantum violations in standard Bell scenarios, that, however, are violated if such communication is allowed.
In order to illustrate the theorem, we present a Bell inequality along with its corresponding CCP, associated to a tripartite Bell scenario with communication related to the \textit{guess-your-neighbour's-input} (GYNI) scenario~\cite{almeida2010guess}. We then proceed to implement this scenario experimentally.
As a second illustration of the Theorem, we introduce the `Svetlichny' scenario \cite{Svet87} in the Supplementary Material.

\medskip

\textbf{Guess-your-neighbour's-input scenario.} The causal structure for this scenario (see \fref{fig:CCP}~(B)), akin to the guess-your-neighbour's-input game \cite{almeida2010guess}, was introduced in \cite{Chaves2017causalhierarchyof}, leading to a new type of multipartite nonlocality.
Classical correlations \eqref{eq:nlhv} are bounded by the inequality \cite{Chaves2017causalhierarchyof}
\begin{eqnarray}
\label{eq:Cyclic}
B_{G}=\sum_{x_1,x_2,x_3=-1}^1Q_{G}(x_1,x_2,x_3)E_{x_1x_2x_3}\leq B^C_{G},
\end{eqnarray}
with $B^C_{G}=6$ and $Q_{G}(x_1,x_2,x_3) = 1 - \frac{(1-x_1)(1-x_2)(1-x_3)}{4}$.
Following the causal structure in \fref{fig:CCP})~B and general prescription of the CCP, party 1 broadcasts a bit $m_1(x_1,x_3,y_1,a_1)$, party 2, $m_2(x_1,x_2,y_2,a_2)$, and party 3, $m_3(x_2,x_3,y_3,a_3)$.

According to the Theorem, the classical probability of success in computing the associated function is  $P^C_{Suc} \leq 7/8=0.875$. 
As shown in \cite{Chaves2017causalhierarchyof}, a quantum violation of this inequality is not possible if party $i$ only has access to the input $x_i$.
If however, the three parties share a GHZ state of the form $\ket{\text{GHZ}}=(\ket{000}+\ket{111})/\sqrt{2}$, and are able to choose their measurements according to the GYNI causal structure depicted in \fref{fig:CCP}~(B), the inequality \eqref{eq:Cyclic} can be violated up to $B_{G} \approx 7.39$. 
This results in a higher probability of success, $P_{Suc} \approx 0.962$, a quantum advantage in this CCP.
As shown in the Supplementary Material, resorting to a generalisation of the NPA hierarchy \cite{navascues2007bounding} (a secondary but still relevant technical contribution of our results), this is the optimal quantum value.

\begin{figure}
 \includegraphics[width=\columnwidth]{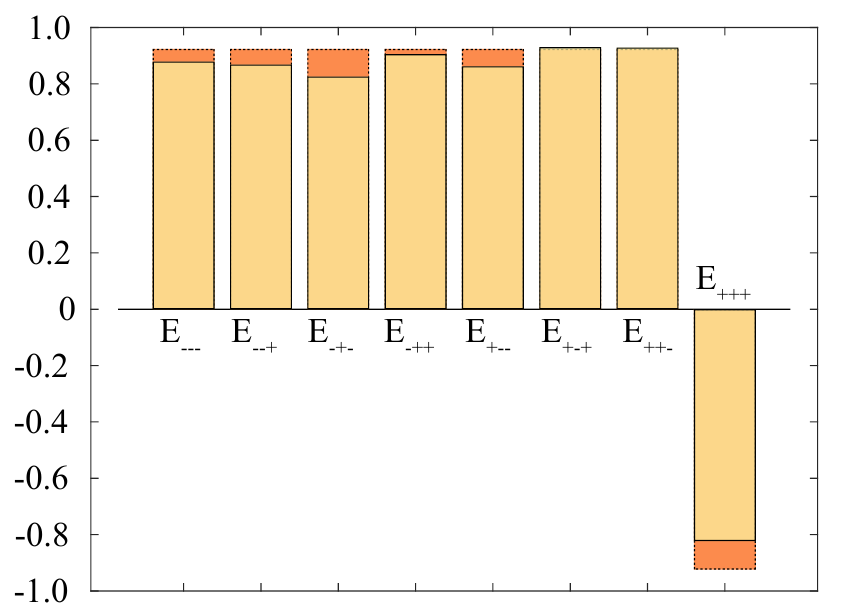}
\caption{\textbf{Experimental results.} Experimentally measured correlation terms belonging to the non-local Bell inequality in the GYNI scenario, \( E_{x_1 x_2 x_3} \) where \(\{x_1, x_2, x_3\} = \pm{1}\). Orange bars represent the theoretical values assuming the optimal measurement settings while yellow bars show the experimentally observed values. We evaluate the inequality and report a correlation value of \(B_G = 7.023 \pm{0.036}\). Errors as estimated by Monte Carlo sampling, using $N=200$ runs and assuming Poissonian statistics, are omitted as they are to small to be visible.}
 \label{fig:inequalityMeasurement}
\end{figure}

\medskip
\textbf{Experimental implementation.} We experimentally investigate the GYNI scenario by producing a tripartite GHZ state encoded in polarisation of telecom-wavelength photons.
We measure the correlation terms defined by inequality \eqref{eq:Cyclic} to demonstrate that the experimentally observed state can violate the inequality, which is both necessary and sufficient for a quantum advantage in the CCP. 
For each correlation term, we implement the optimal measurement settings in Alice, Bob and Charlie's polarisation analysers and record the photon statistics for all \(2^3\) outcomes to evaluate the expectation values as shown in \fref{fig:inequalityMeasurement}. 
See Methods for details on the optimal measurement settings.
From our measurements we obtain a correlation value of \(B_G = 7.023 \pm{0.036}\), representing a violation of the inequality \eqref{eq:Cyclic} by 28 standard deviations with respect to the classical bound of \(B_G^C = 6\).
Using Eq.~\eqref{eq:theorem2}, the observed violation translates to a probability of correctly computing the CCP in this GYNI scenario of \(P_{Suc} = 0.9389 \pm{0.0049}\).

We implement the tripartite GYNI protocol in a faithful round-by-round execution to verify the general connection between CCPs and NLHV models established in this work.
In each round of the protocol, we distribute a GHZ state to three users, and use the NIST randomness beacon~\cite{kelsey2019beacon} to generate randomised input data \((x_1, x_2, x_3, y_1, y_2, y_3)\). 
Each user receives their input bits \(\{x_i, x_j, y_i\}\), updates their polarisation analysers with their respective measurement settings \(M^{(x_i, x_j)}\), and records a single coincidence event to obtain their respective outcomes, \(a_i\), for the round---see Methods for details.
Finally, every user announces their one-bit message, \(m_i = y_i \cdot a_i\), evaluates a guess for the round and compares the joint result with the value of the target function, \(f(\vec{x},\vec{y}) = y_1 y_2 y_3\left(1 - \frac{(1-x_1)(1-x_2)(1-x_3)}{4} \right)\), obtaining a pass/fail for the round.

We perform the protocol for a total of \(10100\) rounds and observe successful outcomes for \(9403\) rounds, corresponding to a probability of success of \(P_{Suc} = 0.9310\).
The experimentally measured probability of success is slightly smaller than the estimated value obtained by the measured violation of the inequality.
This is likely due to drifts in the setup when performing the protocol over the large number of rounds.
For the round-by-round execution we collected two days worth of statistics, whilst the measurements pertaining to the inequality violation were recorded in less than an hour. The majority of data acquisition time in both cases is owed to slow rotation stages that update measurement settings.
Our implementation of the protocol obtained

\section{Discussion}

Generalisations of Bell's theorem to more complex causal networks and in particular those involving communication between the parties are attracting growing attention.
On one side they unveil new \cite{chaves2018quantum,renou2019genuine} and sometimes stronger \cite{Svet87,jones2005extent} kinds of quantum nonlocality.
On the other hand, the practical use of this nonclassicality in the processing of information has so far been limited to specific cases such as device-independent entanglement quantification \cite{moroder2013device}, closing attacks in multipartite cryptographic protocols \cite{moreno2019device} and game theory \cite{moreno2020}.
Here we proposed a general approach by showing that nonlocal hidden variable models introduce a new class of communication complexity problems that contain previous versions \cite{Brukner2004} as special cases.
The Bell inequalities bounding classical correlations in such models can be mapped to the functions to be computed in the associated communication problem.
Further, the violations of these inequalities provide a necessary and sufficient condition for a quantum advantage over the best possible classical protocol.
Our results are theoretically proven in full generality and validated for a specific scenario in an experiment based on a high-fidelity tripartite GHZ state, demonstrating a quantum violation of the Bell inequality \eqref{eq:Cyclic}, akin to the guess-your-neighbour-input scenario~\cite{almeida2010guess,Chaves2017causalhierarchyof}.

Our implementation of the protocol in its entirety---using randomised measurement settings on a round-by-round basis---demonstrates the advantage in a CCP task when using non-locality as a quantum resource.
Future quantum networks will enable the implementation of increasingly sophisticated CCP related tasks over distances, e.g., linking a cluster of network nodes.
Here the entanglement resources are produced in the telecom regime, enabling low-loss transmission in optical fibres connecting network nodes.
Additionally, the investigation of different CCP tasks associated with ``network-friendly'' multipartite entangled states such as graph states would further provide utility in more generalised scenarios.

In spite of the generality of our results a few relevant questions still require further in-depth analysis. As we show here, the violation of a full correlator Bell inequality is a necessary and sufficient condition for a quantum advantage, even in NLHV models related to Bell scenarios with communication.
But for which NLHV models are such violations possible? Initial attempts~\cite{Chaves2017causalhierarchyof} have provided partial answers in the case where the quantum correlations are nonsignalling, that is, the quantum measurements do not make use of the inputs of other parties. The answer to the more general case remains open. Further, we have focused here on full correlator Bell inequalities and it is known that, in the more general case, the violation of Bell inequalities does not necessarily lead to quantum improvements in standard CCPs~\cite{tavakoli_does_2020}. In view of that, analysing under which conditions quantum advantages in generalised CCPs can also be connected with Bell inequalities involving marginals and more measurement outcomes is an interesting question for future research.

\vspace{1em}

\section{Acknowledgements}

We acknowledge the John Templeton Foundation via the Grant Q-CAUSAL No. 61084, the Serrapilheira Institute (Grant No. Serra-1708-15763), the Brazilian National Council for Scientific and Technological Development (CNPq) via the National Institute for Science and Technology on Quantum Information (INCT-IQ), Grants No. 307172/2017-1 and No. 406574/2018-9, the Brazilian agencies MCTIC and MEC, the S\~{a}o Paulo Research Foundation FAPESP (Grant No. 2018/07258-7).
This work was supported by the UK Engineering and Physical Sciences Research Council (Grant Nos. EP/N002962/1 and EP/T001011/1.). FG acknowledges studentship funding from EPSRC under Grant No. EP/L015110/1. 

\vspace{1em}




\section{Methods}

\subsection{Entangled photon source.} 
We employ two parametric down conversion (PDC) sources to create the polarisation-encoded GHZ state.
Each source consists of a \SI{30}{mm} aperiodically poled KTP (aKTP) crystal designed to produce spectrally pure photon pairs at \SI{1550}{nm} in the Type-II configuration~\cite{graffitti2017pure}.
This is achieved through an optimised domain engineering technique, where aperiodic poling achieves a non-linear Phase Matching Function (PMF) that approximates a Gaussian, resulting in near-optimal bi-photon spectral purity~\cite{pickston_optimised_2021}.
This approach allows our photon sources to operate without lossy narrowband filters---the interference filters have nominal full-width-half-maximum bandwith of 8.8nm---allowing higher heralding and collection efficiencies while maintaining high visibility non-classical interference as required for producing multi-photon states efficiently. 
Our crystals are designed for matching a transform-limited Sech-shaped pump spectrum with a pulse duration of \SI{1.3}{ps}.
The domain engineered crystal is embedded in a Sagnac loop which generates polarisation-entanglement between the photon pair. 
A lens with a 50cm nominal focal length is used to focus pump field into each crystal, leading to a source brightness of $\sim$2400 pairs/mW/s and heralding efficiencies of $\sim60\%$.
With 50mW of pump power we witness an interference visibility of $94.2 \pm 1.5\%$ between photons generated from independent sources without any filtering.   
In addition, the picosecond laser is spatially multiplexed attaining \SI{320}{MHz} repetition rate.
This is implemented using two free-space delay loops using 50:50 beamsplitters (BS) and mirrors. 
This allows the peak power per pulse to be reduced to lower the probability of unwanted multi-photon events at the same pump power. 

One photon from each source interferes non-classically on a PBS such that conditional on measuring one photon in each output detector set; Alice (A), Bob (B), Charlie (C), and Trigger (T), the quantum state of the overall four-photon system is,
\begin{equation}
 \ket{ \Psi }_{A,B,C,T} = \frac{ \ket{HVVH} + e^{i \vartheta} \ket{VHHV} }{\sqrt{2}},
\end{equation}
where \(\ket{H}\equiv\ket{0}\) and \(\ket{V} \equiv \ket{1}\) in the logical basis encoding~\cite{Proietti2019ghz}.
We note the phase shift \(\vartheta\) is intrinsic to the optical components in our setup and can be compensated by local operations on any one of the entangled qubits.
Our setup uses standard polarisation measurement analysers---which consists of a QWP, HWP and PBS where the output modes are fibre coupled to SNSPDs---to perform arbitrary projective measurements on each qubit.

To obtain the three-qubit GHZ state for this experiment we project the Trigger photon onto the state, $\ket{\vartheta } \doteq (\ket{H} + e^{-i \vartheta} \ket{V}) / \sqrt{2}$, which ensures entanglement among the remaining photons and simultaneously implements the phase correction.
Detecting a photon after the PBS projects the remaining three photons onto the following state,
\begin{equation}
 \ket{ \Psi }_{A,B,C} = \frac{ \ket{HHH} + \ket{VVV}}{\sqrt{2}},
\end{equation}
up to a local bit-flip which is implemented in the respective user's measurement stage with the use of the polarisation fibre controller.
We perform quantum state tomography and reconstruct the density matrix to characterise the general properties of the GHZ state.
We observe the fidelity to the ideal state to be \(\mathcal{F} = 0.9508 \pm{0.0031}\) and a state purity of \(\mathcal{P} = 0.9255 \pm{0.0058}\).
Uncertainties are reported for one standard deviation and obtained by Monte Carlo sampling using 200 runs, assuming Poissonian statistics.

\subsection{Measurement settings.}
In our experiment the measurement setting in each round for each user is determined by two input bits \(\{x_i, x_j\}\), distributed as per the GYNI scenario.
As such, each user has four possible projective measurement settings, \(M^{(x_i, x_j)}\), given by
\begin{widetext}
\begin{center}
\begin{tabular}{ |l|l|l| } 
 \hline
 \hspace{0.3cm} Alice \hspace{0.3cm} & \hspace{0.4cm} Bob \hspace{0.4cm} & \hspace{0.1cm} Charlie \hspace{0.1cm} \\
 \hline
 \(M_1^{(+, +)} = \frac{\mathbb{I} + \sigma_y}{2}\) & \(M_2^{(+, +)} = \frac{\mathbb{I} + \sigma_x}{2}\) & \(M_3^{(+,+)} = \frac{\mathbb{I} + (-0.38 \sigma_x - 0.92 \sigma_y)}{2}\) \\
 \(M_1^{(+, -)} = \frac{\mathbb{I} - \sigma_y}{2}\) & \(M_2^{(+, -)} = \frac{\mathbb{I}}{2} + \frac{-\sigma_x + \sigma_y}{2 \sqrt{2}}\) & \(M_3^{(+,-)} = \frac{\mathbb{I} + (-0.92 \sigma_x - 0.38 \sigma_y)}{2}\) \\
 \(M_1^{(-, +)} = \frac{\mathbb{I} + \sigma_x}{2}\) & \(M_2^{(-, +)} = \frac{\mathbb{I} + \sigma_y}{2}\) & \(M_3^{(-,+)} = \frac{\mathbb{I} + (-0.38 \sigma_x + 0.92 \sigma_y)}{2}\) \\
 \(M_1^{(-, -)} = \frac{\mathbb{I} + \sigma_y}{2}\) & \(M_2^{(-, -)} = \frac{\mathbb{I}}{2} + \frac{-\sigma_x + \sigma_y}{2 \sqrt{2}}\) & \(M_3^{(-,-)} = \frac{\mathbb{I} + (0.92 \sigma_x - 0.38 \sigma_y)}{2}\) \\
 \hline
\end{tabular}
\end{center}
\end{widetext}

The correlators, \(E_{x_1x_2x_3}=\langle A_{x_1x_3} B_{x_1x_2} C_{x_2x_3} \rangle \), measured in our experiment are expressed in conventional notation in which the subscript indices, denote the local variable \(x_i\) assigned to each user prior to distribution to their neighbour.
As such we make use of the following look-up table to determine the measurements that is performed by each user:

\begin{center}
\begin{tabular}{ |c|c|c|c| } 
 \hline
 Measurement setting & \hspace{0.3cm} Alice \hspace{0.3cm} & \hspace{0.4cm} Bob \hspace{0.4cm} & \hspace{0.1cm} Charlie \hspace{0.1cm} \\
 \hline
 \( E_{+++}  \) & \(M_1^{(+,+ )}\) & \(M_2^{(+, +)}\) & \(M_3^{(+, +)}\) \\
 \(  E_{++-}  \) & \(M_1^{(+, -)}\) & \(M_2^{(-, +)}\) & \(M_3^{(-, +)}\) \\
 \(  E_{+-+}  \) & \(M_1^{(+, +)}\) & \(M_2^{(+, -)}\) & \(M_3^{(+, -)}\) \\
 \( \ E_{+--}  \) & \(M_1^{(+, -)}\) & \(M_2^{(-, -)}\) & \(M_3^{(-, -)}\) \\
 \(  E_{-++} \) & \(M_1^{(-, +)}\) & \(M_2^{(+, +)}\) & \(M_3^{(+, +)}\) \\
 \(  E_{-+-}  \) & \(M_1^{(-, -)}\) & \(M_2^{(-, +)}\) & \(M_3^{(-, +)}\) \\
 \(  E_{--+}  \) & \(M_1^{(-, +)}\) & \(M_2^{(+, -)}\) & \(M_3^{(+, -)}\) \\
 \(  E_{---}  \) & \(M_1^{(-, -)}\) & \(M_2^{(-, -)}\) & \(M_3^{(-, -)}\) \\
 \hline
\end{tabular}
\end{center}

Our measurement apparatus allows us to perform arbitrary projective measurements by using the HWP and QWP to rotate the measurement basis.
Placement of detectors behind both outputs of the PBS enables us to obtain outcomes spanning the full basis set.

\subsection{Calculating correlations.}
In the experiment each user's measurement stage is accompanied with two detectors to measure both outputs of the PBS.
This allows us to directly sample the joint-outcomes for a given basis defined by the measurement settings of each user.
For example to evaluate the correlator \(E_{+++}\) we set the measurement waveplates to implement \(M_1^{(+,+)}, M_2^{(+,+)},\) and \(M_3^{(+,+)}\), for Alice, Bob and Charlie respectively.
We record the three-fold coincidence events according to the outcome detector patterns and evaluate,

\begin{widetext}
    \begin{equation}
        E_{+++}=\langle M_1^{(+,+)} M_2^{(+,+)} M_3^{(+,+)}\rangle = \frac{C_{+++} - C_{++-} - C_{+-+} + C_{+--} - C_{-++} + C_{-+-} + C_{--+} - C_{---}}{C_{+++} + C_{++-} + C_{+-+} + C_{+--} + C_{-++} + C_{-+-} + C_{--+} + C_{---}},
    \end{equation}
\end{widetext}

where \(C_{ijk}\) are the number of coincidences, and indices \(\{i,j,k\} \in \{+,-\}\) denote the outcome for Alice, Bob and Charlie respectively.
Finally, we note that the other correlators are evaluated in the same way following the measurement settings outlined previously.

\subsection{Sketch of the Theorem's Proof.} 
The full proof of \eqref{eq:theorem1} is rather lengthy and presented in the Supplementary Material. Here we focus on \eqref{eq:theorem2}. First notice that full correlators can be written as
\begin{multline}
\label{correlator}
E(x_1,\dots,x_n) = \\S[Q]\left[2P_{x_1,\dots,x_n}\left(\prod_{i=1}^n a_i=S[Q]\right) - 1 \right].
\end{multline}
Inequality \eqref{inequality1} can then be rewritten as
\begin{multline}
\label{inequality2}
\sum_{x_1,\dots,x_n = -1}^1 q^*(x_1,\dots,x_n) \times \\P_{x_1,\dots,x_n}\left(\prod_{i=1}^n a_i=S[Q(x_1,\dots,x_n)]\right)  \leq \frac{1}{2} + \frac{B_n}{2\Gamma}.
\end{multline}
in which $q^*(x_1,\dots,x_n) = \frac{|Q(x_1,\dots,x_n)|}{\Gamma}$. In what follows we will set $q(x_1,\dots,x_n)=q^*(x_1,\dots,x_n)$, thus connecting the probability distribution $q(x_1,...,x_n)$ governing the variables $x_1,...,x_n$ with the coefficients $Q(x_1,...,x_n)$ defining the Bell inequality \eqref{inequality1}.

The protocol proceeds as follows. Each party $i$ chooses a measurement to perform from the set $\left\{ x_{i,j}| j = 1,\dots, l_i\right\}$, obtaining outcome $a_i$.
Each party, then, broadcasts to all other parties the message $m_i=a_i y_i \in\{\pm 1\}$.
In the final step all parties make the same guess about the function $f$ to be computed, given by
\begin{eqnarray}
\label{eq:guess}
G_i = \prod_{j=1}^n m_j=y_1 \dots y_n a_1\dots a_n, \;\;\; \forall i.
\end{eqnarray}
A comparison between the guess of each party \eqref{eq:guess} and the function \eqref{eq:f} to be computed shows that, given a sequence of inputs $x_1\dots x_n, y_1\dots y_n$, the success probability is independent of $y_1$, \dots, $y_n$ and given by $P_{x_1,\dots,x_n}(\prod_{i=1}^n a_i=S[Q(x_1,\dots,x_n)])$.
Hence, since the variables $x_1$, \dots, $x_n$ are sorted according to a distribution $q(x_1,\dots,x_n)$ and variables $y_1$, ..., $y_n$ are uniformly sorted, the final probability of success is
\begin{multline}
\nonumber
P_{Suc} = \sum_{x_1,\dots,x_n = -1}^1 q(x_1,\dots,x_n)\times\\
P_{x_1,\dots,x_n}\left(\prod_{i=1}^n a_i=S[Q(x_{1},...,x_{n})]\right).
\end{multline}
A comparison with \eqref{inequality2} leads directly to \eqref{eq:theorem2}.

\newpage
\onecolumngrid

\section{Supplementary Information}

\renewcommand{\theequation}{S\arabic{equation}}
\renewcommand{\thefigure}{S\arabic{figure}}
\setcounter{equation}{0}
\setcounter{figure}{0}

\section{The Svetlichny scenario}

\begin{figure}[!h]
\includegraphics[width=.35\columnwidth]{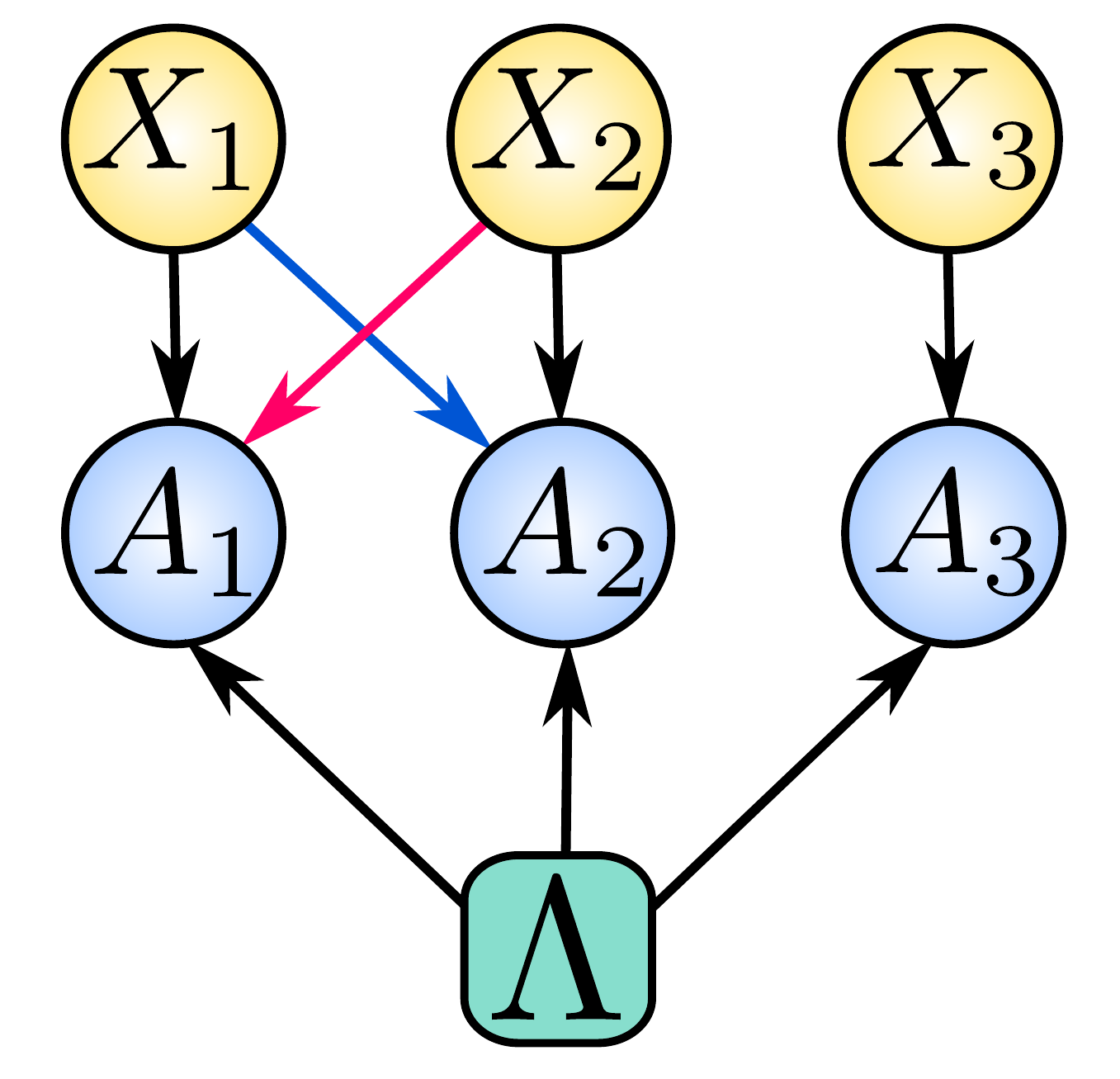}
\caption[]{\label{fig:cstructures_svetlichny} Causal structure of the Svetlichny scenario where any permutation of two parties can exchange their local inputs.}
\end{figure}

Here we provide another example illustrating the main result in our paper, related to the Svetlichny scenario \cite{Svet87} of central importance in the certification of genuine multipartite entanglement. The Svetlichny inequality \cite{Svet87}
\begin{eqnarray}
\label{eq:Svet}
B_{S}=\sum_{x_1,x_2,x_3=-1}^1Q_{S}(x_1,x_2,x_3)E_{x_1,x_2,x_3} \leq B^C_{S},
\end{eqnarray}
with $B^C_{S}$=4 and $Q_{S}(x_1,x_2,x_3) = 1 - \frac{(1-x_1)(1-x_2)(1-x_3)}{4} - \frac{(1+x_1)(1+x_2)(1+x_3)}{4}$,
was the first multi-party Bell inequality related to a stronger notion of nonlocality, since it is based on a NLHV model where any  permutation of two parties can exchange their inputs. Focusing here on the causal structure shown in Fig. \eqref{fig:cstructures_svetlichny} (however, the results hold for any permutation of parties and convex combinations thereof), we see that both parties 1 and 2 receive the inputs $x_1$ and $x_2$, while party 3 only receives $x_3$. Thus, in the execution of the associated CCP, party 1 broadcasts a bit $m_1(x_1,x_2,y_1,a_1)$, party 2 $m_2(x_1,x_2,y_2,a_2)$, and party 3 $m_3(x_3,y_3,a_3)$.

According to the Theorem 1 in the main text, the classical probability of success is $P^C_{Suc} \leq 3/4=0.75$. However, if the three parties share a GHZ state $\ket{GHZ}=(1/\sqrt{2})(\ket{000}+\ket{111})$, the inequality \eqref{eq:Svet} can be violated up to $4\sqrt{2}$, implying a higher probability of success, $P_{Suc}=(1/2)(1+\sqrt{2}/2) \approx 0.853$. As proven bellow, this optimal quantum violation is independent of whether parties 1 and 2 use, each, a single input -- $x_{1}$ and $x_{2}$, respectively --, or use, each, both inputs $x_1$ and $x_2$. In other words, in this scenario the optimal quantum advantage is obtained even if the parties have less accessible information than its classical counterpart.

Numerical results (based on a generalization of the NPA hierarchy \cite{navascues2007bounding}, as described below) show that upper bounds for quantum violations of \ref{eq:Svet} with communication between parties 1 and 2 are very close to $4\sqrt{2}$, with separations smaller than $10^{-7}$.
It is known \cite{Svet87} that this value of $4\sqrt{2}$ can be attained without the use of communication (equivalently, with measurements $M_a^{\vec{x}}$ that don't depend on inputs $x_{i,j}$ for $j>1$).
This can be realized with the GHZ state, $\ket{GHZ} = (1/\sqrt{2})(\ket{000} + \ket{111})$, and with rank-1 projective measurements. Given that the inequality involves only full correlation terms, we may consider only the difference between measurement operators of each party $i$, as $A_i^{x_i} \coloneqq M_{0}^{x_i} - M_{1}^{x_i}$, so that
\begin{equation}
B_S = \sum_{\vec{x}} Q_S(x_1,x_2,x_3)\,\bra{GHZ}\,A^{x_1}_1 \otimes A^{x_2}_2 \otimes A^{x_3}_3\,\ket{GHZ}.
\end{equation}
Measurement operators for each outcome can be retrieved by making $M_{a_i}^{x_i} = (\id + a_i\,A_i^{x_i})/2$, recalling that $a_i \in \{-1,1\}$. Optimal measurement choices are then given by
\begin{align}
A_1^{1} &= -\left(\frac{\sigma_x + \sigma_y}{\sqrt{2}}\right),\\
A_1^{-1} &= \frac{-\sigma_x + \sigma_y}{\sqrt{2}},\\
A_2^{1} &= \sigma_y,\\
A_2^{-1} &= -\sigma_x, \\
A_3^{1} &= \sigma_x,\\
A_3^{-1} &= \sigma_y,
\end{align}
where $\sigma_x$, $\sigma_y$, and $\sigma_z$ are the usual Pauli matrices.

\subsection{Proof of Theorem 1 in tripartite Svetlichny scenario}

The Svetlichny inequality is given by:
\begin{eqnarray}
\label{Svet}
\sum_{x_1,x_2,x_3=-1}^1Q_{S}(x_1,x_2,x_3)E_{x_1x_2x_3} \leq 4,
\end{eqnarray}
which also can be rewritten as follows:
\begin{eqnarray}
\label{Modified Svet}
\frac{1}{8} \sum_{x_1,x_2,x_3=-1}^1P_{x_1x_2x_3}(a_1a_2a_3 = Q_{S}(x_1,x_2,x_3)) \leq \frac{3}{4}.
\end{eqnarray}

Using the strategy described in Methods we get to a success rate of party $i\in\{A,B,C\}$:
\begin{eqnarray}
\mathcal{P}_{i} = \frac{1}{8}\sum_{x_1,x_2,x_3}P(a_1a_2a_3=S[g]|x_1x_2x_3),
\end{eqnarray}
which is clearly more efficient when using correlations that violate \eqref{Modified Svet}.

To complete our demonstration, we must show this strategy to be optimal. With that purpose, let us take a look at the most general messages each party can broadcast:
\begin{eqnarray}
\label{Alice's message svet}
\nonumber
m_A & = & m_A(x_1,y_1,x_2)\\ 
& = & \sum_{j_1,j_2,j_3}A_{j_1j_2j_3}x_1^{j_1}x_2^{j_3}y_1^{j_2},
\end{eqnarray}
\begin{eqnarray}
\label{Bob's message svet}
\nonumber
m_B & = & m_B(x_2,y_2,x_1)\\ 
& = & \sum_{k_1,k_2,k_3}B_{k_1k_2k_3}x_2^{k_1}x_1^{k_3}y_2^{k_2},
\end{eqnarray}
\begin{eqnarray}
\label{Charlie's message svet}
\nonumber
m_C & = & m_C(x_3,y_3)\\ 
& = & \sum_{l_1,l_2}C_{l_1l_2}x_3^{l_1}y_3^{l_2}.
\end{eqnarray}

Notice that in the equations above we are explicitly disregarding the output of each party. The reason is that the optimal strategy in the classical case will be a deterministic one, in which $a_i$ is a deterministic function of its inputs. For instance, if $a_1(x_1,x_3)$ is a deterministic function, then  $m^{\prime}_A(x_1,y_1,x_3,a_1(x_1,x_3))= m_A(x_1,y_1,x_3)$. In all proofs that follow this will be the case.

Thus, the most general guess Alice may provide is:
\begin{eqnarray}
\label{Alice's Guess svet}
\nonumber
G_A & = & G_A(x_1,y_1,x_2,m_B,m_C)\\
& = & \sum_{q_1,q_2,q_3,q_4,q_5}G_{q_1q_2q_3q_4q_5}x_1^{q_1}y_1^{q_2}x_2^{q_3}m_B^{q_4}m_C^{q_5}.
\end{eqnarray}

Recall that the distributed function that the parties should be able to compute is
\begin{eqnarray}
f(x_{1},x_{2},x_{3},y_{1},y_{2},y_{3}) = y_{1}y_{2}y_{3}Q_{S}(x_{1},x_{2},_{3}).
\end{eqnarray}
The fidelity between the function and Alice's guess, $(f,G_A)$, is defined as:
\begin{eqnarray}
\nonumber
(f,G_A) & = & \frac{1}{64}\sum_{x_1,x_2,x_3,y_1,y_2,y_3=-1}^1 f(x_1,y_1,x_2,y_2,x_3,y_3)G_A(x_1,y_1,x_2,m_B,m_C)\\ \nonumber
& = & \frac{1}{64}\sum_{x_1,x_2,x_3,y_1,y_2,y_3=-1}^1 y_1y_2y_3Q_{S}(x_1,x_2,x_3)\sum_{q_1,q_2,q_3,q_4,q_5}G_{q_1q_2q_3q_4q_5}x_1^{q_1}y_1^{q_2}x_2^{q_3}m_B^{q_4}m_C^{q_5}\\ \nonumber
& = & \frac{1}{32}\sum_{x_1,x_2,x_3=-1}^1 Q_{S}(x_1,x_2,x_3)\left(\sum_{q_1,q_3}G_{q_11q_311}x_1^{q_1}x_2^{q_3}\right)\left(\sum_{y_2=-1}^1y_2m_B\right)\left(\sum_{y_3=-1}^1y_3m_C\right)\\ \nonumber
& = & \frac{1}{8}\sum_{x_1,x_2,x_3=-1}^1 Q_{S}(x_1,x_2,x_3)\left(\sum_{q_1,q_3}G_{q_11q_311}x_1^{q_1}x_2^{q_3}\right)\left(\sum_{k_1,k_3}B_{k_11k_3}x_2^{k_1}x_1^{k_3}\right)\left(\sum_{l_1}C_{l_11}x_3^{l_1}\right)
\end{eqnarray}

Notice that from equation \eqref{Bob's message svet}, we have:
\begin{eqnarray}
\sum_{y_2=-1}^1y_2m_B(x_2,y_2,x_1) = 2\sum_{k_1,k_3}B_{k_11k_3}x_2^{k_1}x_1^{k_3},
\end{eqnarray}
and since $m_B(x_2,y_2,x_1)\in\{-1,1\}$, we must have:
\begin{eqnarray}
\left|\sum_{k_1,k_3}B_{k_11k_3}x_2^{k_1}x_1^{k_3}\right|\leq 1.
\end{eqnarray}
The same argument can be applied on equation \eqref{Charlie's message svet} to obtain:
\begin{eqnarray}
\left|\sum_{l_1}C_{l_11}x_3^{k_1}\right|\leq 1,
\end{eqnarray}
and to equation \eqref{Alice's Guess svet} to obtain:
\begin{eqnarray}
\left|\sum_{q_1,q_3}G_{q_11q_311}x_1^{q_1}x_2^{q_3}\right|\leq 1.
\end{eqnarray}
Using these relations, we can define:
\begin{eqnarray}
\left\{\begin{array}{ll}
      E_{x_1x_2} & = \sum_{q_1,q_3}G_{q_11q_311}x_1^{q_1}x_2^{q_3},\\
      E'_{x_1x_2} & = \sum_{k_1,k_3}B_{k_11k_3}x_2^{k_1}x_1^{k_3},\\
      E_{x_3} & = \sum_{l_1}C_{l_11}x_3^{k_1}.
\end{array}\right.
\end{eqnarray}
This leads to the following expression:
\begin{eqnarray}
\nonumber
(f,G_A) & = & \frac{1}{8}\sum_{x_1,x_2,x_3=-1}^1 Q_{S}(x_1,x_2,x_3)E_{x_1x_2}E'_{x_1x_2}E_{x_2x_3},
\end{eqnarray}
which, from inequality \eqref{Svet}, is clearly bounded by:
\begin{eqnarray}
(f,G_A) & \leq & \frac{1}{2}
\end{eqnarray}

Once the success probability reads:
\begin{eqnarray}
\mathcal{P}_{A} = \frac{1}{2} + \frac{(f,G_A)}{2}
\end{eqnarray}
We have that:
\begin{eqnarray}
\mathcal{P}_{A} \leq \frac{3}{4}.
\end{eqnarray}
The same holds for Bob, given the symmetry of the problem.

On the other hand, the most general guess Charlie may provide is:
\begin{eqnarray}
\label{Charlies's Guess svet}
\nonumber
G_C & = & G_C(x_3,y_3,m_B,m_A)\\
& = & \sum_{q_1,q_2,q_3,q_4}G_{q_1q_2q_3q_4}x_3^{q_1}y_3^{q_2}m_B^{q_3}m_A^{q_4}.
\end{eqnarray}

The fidelity between function $f(x_{1},x_{2},x_{3},y_{1},y_{2},y_{3})$ and Charlie's guess, $(f,G_C)$, reads:
\begin{eqnarray}
\nonumber
(f,G_C) & = & \frac{1}{64}\sum_{x_1,x_2,x_3,y_1,y_2,y_3=-1}^1 f(x_1,y_1,x_2,y_2,x_3,y_3)G_A(x_3,y_3,m_B,m_A)\\ \nonumber
& = & \frac{1}{64}\sum_{x_1,x_2,x_3,y_1,y_2,y_3=-1}^1 y_1y_2y_3Q_{S}(x_1,x_2,x_3)\sum_{q_1,q_2,q_3,q_4}G_{q_1q_2q_3q_4}x_3^{q_1}y_3^{q_2}m_B^{q_3}m_A^{q_4}\\ \nonumber
& = & \frac{1}{32}\sum_{x_1,x_2,x_3=-1}^1 Q_{S}(x_1,x_2,x_3)\left(\sum_{q_1}G_{q_1111}x_1^{q_1}\right)\left(\sum_{y_2=-1}^1y_2m_B\right)\left(\sum_{y_1=-1}^1y_1m_A\right)\\ \nonumber
& = & \frac{1}{8}\sum_{x_1,x_2,x_3=-1}^1 Q_{S}(x_1,x_2,x_3)\left(\sum_{q_1}G_{q_1111}x_3^{q_1}\right)\left(\sum_{k_1,k_3}B_{k_11k_3}x_2^{k_1}x_1^{k_3}\right)\left(\sum_{j_1,j_3}A_{j_11j_3}x_1^{j_1}x_2^{j_3}\right).
\end{eqnarray}
Notice that, from equation \eqref{Bob's message svet}, we have:
\begin{eqnarray}
\sum_{y_2=-1}^1y_2m_B(x_2,y_2,x_1) = 2\sum_{k_1,k_3}B_{k_11k_3}x_2^{k_1}x_1^{k_3},
\end{eqnarray}
and since $m_B(x_2,y_2,x_1)\in\{-1,1\}$, we must have:
\begin{eqnarray}
\left|\sum_{k_1,k_3}B_{k_11k_3}x_2^{k_1}x_1^{k_3}\right|\leq 1.
\end{eqnarray}
The same argument can be applied on equation \eqref{Alice's message svet} to obtain:
\begin{eqnarray}
\left|\sum_{j_1,j_3}A_{j_11j_3}x_1^{j_1}x_2^{j_3}\right|\leq 1,
\end{eqnarray}
and to equation \eqref{Charlies's Guess svet} to obtain:
\begin{eqnarray}
\left|\sum_{q_1}G_{q_1111}x_3^{q_1}\right|\leq 1.
\end{eqnarray}
Using these relations, we can define:
\begin{eqnarray}
\left\{\begin{array}{ll}
      E_{x_1x_2} & = \sum_{j_1,j_3}A_{j_11j_3}x_1^{j_1}x_2^{j_3},\\
      E'_{x_1x_2} & = \sum_{k_1,k_3}B_{k_11k_3}x_2^{k_1}x_1^{k_3},\\
      E_{x_3} & = \sum_{q_1}G_{q_1111}x_3^{q_1}.
\end{array}\right.
\end{eqnarray}
These lead to:
\begin{eqnarray}
\nonumber
(f,G_C) & = & \frac{1}{8}\sum_{x_1,x_2,x_3=-1}^1 Q_{S}(x_1,x_2,x_3)E_{x_1x_2}E'_{x_1x_2}E_{x_3},
\end{eqnarray}
which, from inequality \eqref{Svet}, is clearly bounded by:
\begin{eqnarray}
(f,G_C) & \leq & \frac{1}{2}
\end{eqnarray}

Once the success probability reads:
\begin{eqnarray}
\mathcal{P}_{C} = \frac{1}{2} + \frac{(f,G_C)}{2},
\end{eqnarray}
we have that:
\begin{eqnarray}
\mathcal{P}_{C} \leq \frac{3}{4}.
\end{eqnarray}
This shows, once again, that Theorem 1 holds.

\section{Guess-your-neighbour-input scenario}
Performing a brute force optimization over all pure qubit states and projective measurements we have found that that maximum violation of the inequality (12) in the main text is given by $S_{G}\approx 7.391$. As discussed below, this value is still pretty close to the numerical value obtained by running an extension of the NPA hierarchy \cite{navascues2007bounding} for a Bell scenario with communication. In this case, the optimization is performed over all possible quantum states and measurements but in general it will only provide an upper bound for the maximum quantum value.

Regarding the maximum quantum violation with qubits. The projective measurement for each of the three parties can be written as 
\begin{eqnarray}
M_{a_1|x_1,x_3} = \frac{\id}{2}+ a_1\frac{\vec{r}_{A_1}(x_1,x_3)\cdot\vec{\sigma}}{2}\\
M_{a_2|x_2,x_1} = \frac{\id}{2}+ a_2\frac{\vec{r}_{A_2}(x_2,x_1)\cdot\vec{\sigma}}{2}\\
M_{a_3|x_3,x_2} = \frac{\id}{2}+ a_3\frac{\vec{r}_{A_3}(x_3,x_2)\cdot\vec{\sigma}}{2}
\end{eqnarray}{}
where $\vec\sigma = (\sigma_x,\sigma_y,\sigma_z)$ are the Pauli matrices.

By performing a brute-force numerical optimization the best state is a GHZ state and the optimal measurements are given by
\begin{eqnarray}
M_{a_1|1,1} &=& \frac{\id+ a_1\sigma_y}{2} \nonumber \\
M_{a_1|1,-1} &=& \frac{\id+ a_1\sigma_y}{2} \nonumber\\
M_{a_1|-1,1} &=& \frac{\id+ a_1\sigma_x}{2}\nonumber\\
M_{a_1|-1,-1} &=& \frac{\id+ a_1\sigma_y}{2}\nonumber\\
M_{a_2|1,1} &=& \frac{\id+ a_2\sigma_x}{2}\nonumber\\
M_{a_2|1,-1} &=& \frac{\id}{2}+ a_2\frac{(-\sigma_x + \sigma_y)}{2\sqrt{2}}\nonumber\\
M_{a_2|-1,1} &=& \frac{\id+ a_2\sigma_y}{2}\nonumber\\
M_{a_2|-1,-1} &=& \frac{\id}{2}+ a_2\frac{(-\sigma_x + \sigma_y)}{2\sqrt{2}}\nonumber\\
M_{a_3|1,1} &=& \frac{\id+ a_3(-0.38\sigma_x - 0.92\sigma_y)}{2}\nonumber\\
M_{a_3|1,-1} &=& \frac{\id+ a_3(-0.92\sigma_x - 0.38\sigma_y)}{2}\nonumber\\
M_{a_3|-1,1} &=& \frac{\id+ a_3(-0.38\sigma_x + 0.92\sigma_y)}{2}\nonumber\\
M_{a_3|-1,-1} &=& \frac{\id+ a_3(0.92\sigma_x - 0.38\sigma_y)}{2}\nonumber\\
\end{eqnarray}

\subsection{Proof of Theorem 1 in tripartite Guess-your-neighbour-inputs scenario}
\label{intro}
In this scenario we have that: Alice receives $\{x_1,y_1,x_3\}$, Bob receives $\{x_2,y_2,x_1\}$, and Charlie receives $\{x_3,y_3,x_2\}$.

The goal is to evaluate some function $f$ specified below.

\begin{eqnarray}
\nonumber
f & = & f(x_1,y_1,x_2,y_2,x_3,y_3)\\ 
& = & y_1y_2y_3S[Q_{G}(x_1,x_2,x_3)],
\end{eqnarray}
in which:
\begin{eqnarray}
Q_{G}(x_1,x_2,x_3) = 1 - \frac{(1-x_1)(1-x_2)(1-x_3)}{4}
\end{eqnarray}
and $S[.]$ is the sign function (in this case $S[Q_{G}]=Q_{G}$).

An inequality bounding the set of classical behaviours in the three partite cyclic scenario can be expressed as:
\begin{eqnarray}
\label{Main inequality}
\sum_{x_1,x_2,x_3=-1}^1Q_{G}(x_1,x_2,x_3)E_{x_1x_2x_3}\leq 6
\end{eqnarray}

From the definition of $E_{x_1x_2x_3}$, we get to:
\begin{eqnarray}
\label{Modified Inequality}
\frac{1}{8}\sum_{x_1,x_2,x_3}P_{x_1x_2x_3}(a_1a_2a_3=S[Q_G]) \leq \frac{7}{8}
\end{eqnarray}

The strategy described in Methods leads to a success rate for an arbitrary part, say Alice, of:
\begin{eqnarray}
\mathcal{P}_A = \frac{1}{8}\sum_{x_1,x_2,x_3}P(a_1a_2a_3=S[Q_G]|x_1x_2x_3).
\end{eqnarray}
 This proves that in this scenario the violation of this inequality \ref{Modified Inequality} implies an improvement on the success rate.

To answer if the above strategy is the optimal one as called in theorem 1 of the main text, which will imply that the success probability can be improved further than $\frac{7}{8}$ if and only if inequality \ref{Modified Inequality} is violated, we look first to the most general messages each party can broadcast:
\begin{eqnarray}
\label{Alice's message}
\nonumber
m_A & = & m_A(x_1,y_1,x_3)\\
& = & \sum_{j_1,j_2,j_3}A_{j_1j_2j_3}x_1^{j_1}x_3^{j_3}y_1^{j_2}
\end{eqnarray}
\begin{eqnarray}
\label{Bob's message}
\nonumber
m_B & = & m_B(x_2,y_2,x_1)\\
& = & \sum_{k_1,k_2,k_3}B_{k_1k_2k_3}x_2^{k_1}x_1^{k_3}y_2^{k_2}
\end{eqnarray}
\begin{eqnarray}
\label{Charlie's message}
\nonumber
m_C & = & m_C(x_3,y_3,x_2)\\
& = & \sum_{l_1,l_2,l_3}C_{l_1l_2l_3}x_3^{l_1}x_2^{l_3}y_3^{l_2}
\end{eqnarray}

Now we direct our attention to the most general guess one party, say Alice, can provide:
\begin{eqnarray}
\label{Alice's Guess}
\nonumber
G_A & = & G_A(x_1,y_1,x_3,m_B,m_C)\\
& = & \sum_{q_1,q_2,q_3,q_4,q_5}G_{q_1q_2q_3q_4q_5}x_1^{q_1}y_1^{q_2}x_3^{q_3}m_B^{q_4}m_C^{q_5}
\end{eqnarray}

The fidelity $(f,G_A)$ reads:
\begin{eqnarray}
\nonumber
(f,G_A) & = & \frac{1}{64}\sum_{x_1,x_2,x_3,y_1,y_2,y_3=-1}^1 f(x_1,y_1,x_2,y_2,x_3,y_3)G_A(x_1,y_1,x_3,m_B,m_C)\\ \nonumber
& = & \frac{1}{64}\sum_{x_1,x_2,x_3,y_1,y_2,y_3=-1}^1 y_1y_2y_3Q_G(x_1,x_2,x_3)\sum_{q_1,q_2,q_3,q_4,q_5}G_{q_1q_2q_3q_4q_5}x_1^{q_1}y_1^{q_2}x_3^{q_3}m_B^{q_4}m_C^{q_5}\\ \nonumber
& = & \frac{1}{32}\sum_{x_1,x_2,x_3=-1}^1 Q_{G}(x_1,x_2,x_3)\left(\sum_{q_1,q_3}G_{q_11q_311}x_1^{q_1}x_3^{q_3}\right)\left(\sum_{y_2=-1}^1y_2m_B\right)\left(\sum_{y_3=-1}^1y_3m_C\right)\\ \nonumber
& = & \frac{1}{8}\sum_{x_1,x_2,x_3=-1}^1 Q_{G}(x_1,x_2,x_3)\left(\sum_{q_1,q_3}G_{q_11q_311}x_1^{q_1}x_3^{q_3}\right)\left(\sum_{k_1,k_3}B_{k_11k_3}x_2^{k_1}x_1^{k_3}\right)\left(\sum_{l_1,l_3}C_{l_11l_3}x_3^{l_1}x_2^{l_3}\right)
\end{eqnarray}

Notice that from equation \ref{Bob's message}, we have:
\begin{eqnarray}
\sum_{y_2=-1}^1y_2m_B(x_2,y_2,x_1) = 2\sum_{k_1,k_3}B_{k_11k_3}x_2^{k_1}x_1^{k_3},
\end{eqnarray}
and once $m_B(x_2,y_2,x_1)\in\{-1,1\}$ we must have:
\begin{eqnarray}
\left|\sum_{k_1,k_3}B_{k_11k_3}x_2^{k_1}x_1^{k_3}\right|\leq 1.
\end{eqnarray}

The same argument can be applied on equation \ref{Charlie's message} to obtain:
\begin{eqnarray}
\left|\sum_{l_1,l_3}C_{l_11l_3}x_3^{k_1}x_2^{k_3}\right|\leq 1,
\end{eqnarray}
and to equation \ref{Alice's Guess} to obtain:
\begin{eqnarray}
\left|\sum_{q_1,q_3}G_{q_11q_311}x_1^{q_1}x_3^{q_3}\right|\leq 1.
\end{eqnarray}

Using these relations, we can define:
\begin{eqnarray}
\left\{\begin{array}{ll}
     E_{x_1x_3} & = \sum_{q_1,q_3}G_{q_11q_311}x_1^{q_1}x_3^{q_3}\\
     E_{x_1x_2} & = \sum_{k_1,k_3}B_{k_11k_3}x_2^{k_1}x_1^{k_3}\\
     E_{x_2x_3} & = \sum_{l_1,l_3}C_{l_11l_3}x_3^{k_1}x_2^{k_3}
\end{array}\right.
\end{eqnarray}

Leading to:
\begin{eqnarray}
\nonumber
(f,G_A) & = & \frac{1}{8}\sum_{x_1,x_2,x_3=-1}^1 Q_{G}(x_1,x_2,x_3)E_{x_1x_3}E_{x_1x_2}E_{x_2x_3}
\end{eqnarray}
Which, given inequality \ref{Main inequality}, clearly is bounded to:
\begin{eqnarray}
(f,G_A) & \leq & \frac{6}{8}
\end{eqnarray}

Once the success probability reads:
\begin{eqnarray}
\mathcal{P}_{A} = \frac{1}{2} + \frac{(f,G_A)}{2}
\end{eqnarray}

We have that:
\begin{eqnarray}
\mathcal{P}_{A} \leq \frac{7}{8}
\end{eqnarray}
proving that theorem 1 holds.

Notice that this result holds for every party given the symmetry of the problem.

\section{Generalization of the NPA hierarchy to scenarios with communication}

In this section we show how to estimate an upper bound on the success of quantum strategies. Since a direct computation of the best quantum strategy is a difficult problem, we resort to successive approximations of the set of quantum probabilities by a Navascues-Pironio-Acin (NPA) hierarchy of supersets \cite{navascues2007bounding}, with supersets of higher levels in the hierarchy contained in all sets of lower level, in a way that the sequence converges exactly to the quantum set as the level goes to infinity.

\begin{figure}[!h]
\includegraphics[width=.5\columnwidth]{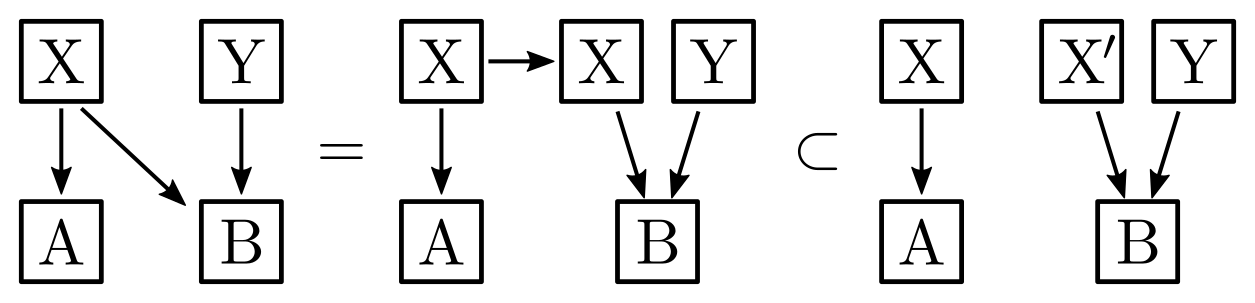}
\caption{Adaptation of the NPA method to signaling scenarios. We find an augmented, nonsignaling scenario that includes the signaling one as a particular case. Nonlocal influences in the original model, such as $X \rightarrow B$ in the figure, become then mediated by local variables ($X^\prime$ in the example), which are considered as independent variables. The original problem is then obtained when the particular choice $x^\prime=x$ is made.}
\label{fig:app_NPAScheme}
\end{figure}

To employ this technique in a scenario that involves signalling among the parties, we assume that a genuine quantum realization of any given probability distribution implies the existence of measurement operators $M^{x,x'}_a$ for each possible combination of local and nonlocal inputs (represented by $x$ and $x'$, respectively). Effectively, this means that we take the signalling scenario as a particular case of a larger nonsignaling scenario, where local-input alphabets are augmented to incorporate the influence of communicating parties. This notion is illustrated in Fig. \ref{fig:app_NPAScheme}. 

In particular, for instance, a quantum realization for a distribution compatible with the guess-your-neighbours-input scenario, shown in Fig. 1b of the main text, is given by positive semidefinite operators $M^{x,u}_a, M^{y,v}_b, M^{z,w}_c$ and a density matrix $\rho$ such that
\begin{equation}
P(abc\vert (x,u)\,(y,v)\,(z,w)) = \Tr\left[\left(M^{x, u}_a \otimes M^{y, v}_b \otimes M^{z, w}_c\right)\, \rho\right],
\end{equation}
and $\sum_a M^{x,u}_a = \id$ (and similarly for the other parties), which in turn realize the observed distribution when the identifications $u=z,\, v=x,\, w=y$ are executed, so that $P(abc\vert xyz) = P(abc\vert (x,z)\,(y,x)\,(z,y))$. 

We then solve the approximate compatibility problem with a standard semidefinite program (SDP). The problem consists of finding a truncated, positive semidefinite moment matrix compatible with the augmented probability, with the corresponding internal constraints and the additional constraint that the reduced distribution matches the observed one. Formally, the SDP is given by
\begin{subequations}
\label{eq:app_NPA_SDP}
\begin{align}
\textrm{Given} &\quad Q_{G} \\
\maximize_{P, \mathcal{M}} &\quad \sum_{x_1,x_2,x_3=-1}^1 Q_{G}(x_1,x_2,x_3)\left[P_{\textrm{sig}}(\prod_i a_i=1\vert  x_1x_2x_3) - P_{\textrm{sig}}(\prod_i a_i=-1\vert x_1x_2x_3)\right], \\
\st &\quad  P_{\textrm{sig}}(a_1a_2a_3\vert x_1x_2x_3) = P(a_1a_2a_3\vert (x_1,x_3)\,(x_2,x_1)\,(x_3,x_2))\\
&\quad P(a_1a_2a_3\vert (x,u)\,(y,v)\,(z,w)) \geq 0 \quad \forall\, a_1,\ a_2,\ a_3,\ x,\ u,\ y,\ v,\ z,\ w \\
&\quad \mathcal{M} \geq 0,\quad \mathcal{M}^\dagger = \mathcal{M},\\
&\quad \mathcal{M} \in \mathrm{NPA}_k
\end{align}
\end{subequations}
where the optimization variable $P$ corresponds to a distribution in the augmented scenario; $P_{\textrm{sig}}$ is the distribution of interest in the signaling scenario. $\mathcal{M}$ is a moment matrix, truncated according to the level of approximation desired (set as $k$ in the program description). 

A valid moment matrix is one with entries $\mathcal{M}_{ij}$ compatible with $\Tr\left[A_i A_j^\dagger\, \rho \right]$, for some density matrix $\rho$ and $A_i$ operators in a set $\mathcal{S} = \{\id\} \bigcup \mathcal{S}_1 \bigcup \ldots \bigcup \mathcal{S}_k$, with $\mathcal{S}_n = \{AB\,\vert\ A \in \mathcal{S}_1,\, B \in \mathcal{S}_{n-1}\}$ for $n > 1$, and $\mathcal{S}_1$ a nonempty set of operators, chosen, in our case, as the set of POVMs $M^{x,u}_a,\,M^{y,v}_b,\,M^{z,w}_c$, with extra terms corresponding to combinations of three of these operators. Constraints over $\mathcal{M}$ correspond to these compatibility conditions. In particular, pertinence to $NPA_k$ corresponds to linear equality constraints among the entries of $\mathcal{M}$ and between entries and elements of the distribution $P$ imposed by the expected structure.

In this way, we have obtained a violation of $S_G \approx 7.393$ using moments generated by the second level in the hierarchy with extra measurements combining extra terms $ABC + AAB + AAC + BCB$. Considering the Svetlichny's scenario (Fig. \ref{fig:cstructures_svetlichny}) instead, and using inequality \eqref{Svet} of the main text, we have obtained $S_S \approx 5.6568$, practically coinciding with the quantum bound $4\sqrt{2}$, already for the second level of the hierarchy with no extra terms. We used Peter Wittek's \emph{ncpol2sdpa} library for Python \cite{Wittek2015} to obtain the moment matrix structure for the SDP and we solved it using \emph{MOSEK} solver \cite{mosek}.

\section{Proof of Theorem 1 in the general case}

Here, we consider the general case in which a source distributes $2n$ bits $\{x_j,y_j\}_{j=1,...,n}$ among $n$ separated local parties. Party $i$ receives the bits $\{x_i,y_i\}$ plus $l_i-1$ other bits from the set $\{x_j\}_{j\neq i}$. For convenience, we adopt a relabelling of the inputs communicated to party $i$ with respect to party $i$ as follows: the new label consists of two sub-indexes the first one stand for the party we are taking as reference and the second is the relabel itself which is take to be sequential, for instance all variables $x_k$ that are communicated to party $i$ are relabeled as $x_{i,j}$ with $i\in{1,...,l_i}$ (in particular we take $x_{i,1}=x_i$). Furthermore, using the same notation, we relabel the variables that are not communicated to party $i$ with a tilde: the subset of $\{x_j,y_j\}_{j=1,...,n}$ of variables which are not communicated to party $i$ is specified by a tilde sign, forming a set $\{\tilde{x}_{i,j}\}_{i=1,...,n-l_i}$.

Each party $j$ broadcasts a message to the other parties, the message $j$ sends to party $i$ will be referred as message $m_{i,j}$.

The goal for each party is the evaluation of function $f$:
\begin{eqnarray}
f(x_{i,1},...,x_{i,l_i},\tilde{x}_{i,1},...,\tilde{x}_{i,n-l_i},y_{i,1},...,y_{i,n})=y_{i,1}...y_{i,n}S[Q(x_{i,1},...,x_{i,l_i},\tilde{x}_{i,1},...,\tilde{x}_{i,n-l_i})]
\end{eqnarray}

\subsection{Inequality and strategy}
The following equation represents an inequality bounding the set of classical behaviours in an N-partite Bell scenario with arbitrary relaxations on the local assumption:
\begin{eqnarray}
\label{Main inequality (n parts)}
 B_n = \sum_{\substack{x_{i,1},...,x_{i,l_i}=-1\\ \tilde{x}_{i,1},...,\tilde{x}_{i,n-l_i}=-1}}^1Q(x_{i,1},...,x_{i,l_i},\tilde{x}_{i,1},...,\tilde{x}_{i,n-l_i})E_{x_{i,1}...x_{i,l_i}\tilde{x}_{i,1}...\tilde{x}_{i,n-l_i}}\leq B_n^C
\end{eqnarray}
in which $E_{x_{i,1}...x_{i,l_i}\tilde{x}_{i,1}...\tilde{x}_{i,n-l_i}}$ is the correlation function:
\begin{eqnarray}
\label{Correlation definition (n parts)}
E_{x_{i,1}...x_{i,l_i}\tilde{x}_{i,1}...\tilde{x}_{i,n-l_i}} = S[h]\left(2P_{x_{i,1}...x_{i,l_i}\tilde{x}_{i,1}...\tilde{x}_{i,n-l_i}}\left(\prod_{j=1}^na_j=S[h]\right) - 1\right)
\end{eqnarray}
for any real number $h$.

Setting $h=Q(x_{i,1},...,x_{i,l_i},\tilde{x}_{i,1},...,\tilde{x}_{i,n-l_i})$ and using equation \eqref{Correlation definition (n parts)} on \ref{Main inequality (n parts)} we get to:
\begin{eqnarray}
\label{Modified Inequality (n parts)}
\sum_{\substack{x_{i,1},...,x_{i,l_i}=-1\\ \tilde{x}_{i,1},...,\tilde{x}_{i,n-l_i}=-1}}^1q(x_{i,1},...,x_{i,l_i},\tilde{x}_{i,1},...,\tilde{x}_{i,n-l_i})P_{x_{i,1}...x_{i,l_i}\tilde{x}_{i,1}...\tilde{x}_{i,n-l_i}}\left(\prod_{j=1}^na_j=S[Q]\right) \leq \frac{1}{2} + \frac{B_n^C}{2\Gamma};
\end{eqnarray}
in which:
\begin{eqnarray}
\label{Q (N parts}
q(x_{i,1},...,x_{i,l_i},\tilde{x}_{i,1},...,\tilde{x}_{i,n-l_i})= \frac{|Q(x_{i,1},...,x_{i,l_i},\tilde{x}_{i,1},...,\tilde{x}_{i,n-l_i})|}{\Gamma},
\end{eqnarray}
and:
\begin{eqnarray}
\Gamma = \sum_{\substack{x_{i,1},...,x_{i,l_i}=-1\\ \tilde{x}_{i,1},...,\tilde{x}_{i,n-l_i}=-1}}^1|Q(x_{i,1},...,x_{i,l_i},\tilde{x}_{i,1},...,\tilde{x}_{i,n-l_i})|.
\end{eqnarray}

Notice that the function $q(x_{i,1},...,x_{i,l_i},\tilde{x}_{i,1},...,\tilde{x}_{i,n-l_i})$ has the properties of a distribution:
\begin{eqnarray}
\left\{\begin{array}{l}
      0\leq q(x_{i,1},...,x_{i,l_i},\tilde{x}_{i,1},...,\tilde{x}_{i,n-l_i})\leq 1\\
      \sum_{\substack{x_{i,1},...,x_{i,l_i}=-1\\ \tilde{x}_{i,1},...,\tilde{x}_{i,n-l_i}=-1}}^1 q(x_{i,1},...,x_{i,l_i},\tilde{x}_{i,1},...,\tilde{x}_{i,n-l_i})=1
\end{array}\right.
\end{eqnarray}

Following the strategy proposed in the main text party $i$ has a success rate of:
\begin{eqnarray}
\mathcal{P}_{i} = \sum_{\substack{x_{i,1},...,x_{i,l_i}=-1\\ \tilde{x}_{i,1},...,\tilde{x}_{i,n-l_i}=-1}}^1q(x_{i,1},...,x_{i,l_i},\tilde{x}_{i,1},...,\tilde{x}_{i,n-l_i})P_{x_{i,1}...x_{i,l_i}\tilde{x}_{i,1}...\tilde{x}_{i,n-l_i}}\left(\prod_{j=1}^na_j=S[Q]\right)
\end{eqnarray}
which is clearly more efficient when using correlations that violate \ref{Modified Inequality (n parts)}.

\subsection{Is it optimal?}
The most general message party $j$ can send to party $i$ is:
\begin{eqnarray}
\label{Message N parts}
m_{i,j} = \sum_{r_{0},...,r_{l_j}=0}^1B^{(i,j)}_{r_{0}...r_{l_j}}y_{i,j}^{r_{0}}\prod_{u=1}^{l_j}x_{j,u}^{r_{u}}
\end{eqnarray}

The most general guess party $i$ can provide is:
\begin{eqnarray}
\label{Guess N parts}
G_i(x_{i,1},...,x_{i,l_i},y_{i},m_{i,2},...,m_{i,n}) = \sum_{\substack{s_0,...,s_{l_i}=0\\ t_2,...,t_n=0}}^1G_{s_0...s_{l_i}t_2...t_n} y_{i,1}^{s_0}\left(\prod_{v=1}^{l_i}x_{i,v}^{s_v}\right)\left(\prod_{w=2}^n m_{i,w}^{t_w}\right)
\end{eqnarray}

The success probability can be expressed in terms of the fidelity $(f,G_i)$:
\begin{eqnarray}
\mathcal{P}_{i} = \frac{1}{2} + \frac{(f,G_i)}{2}
\end{eqnarray}

The most general fidelity reads:
\begin{eqnarray}
\nonumber
(f,G_i) & = & \frac{1}{2^n}\sum_{\substack{x_{i,1},...,x_{i,l_i}=-1\\ \tilde{x}_{i,1},...,\tilde{x}_{i,n-l_i}=-1\\y_{i,1},...,y_{i,n}=-1}}^1q(x_{i,1},...,x_{i,l_i},\tilde{x}_{i,1},...,\tilde{x}_{i,n-l_i})fG_i
\\ 
\nonumber
& = & \frac{1}{2^n\Gamma}\sum_{\substack{x_{i,1},...,x_{i,l_i}=-1\\ \tilde{x}_{i,1},...,\tilde{x}_{i,n-l_i}=-1\\y_{i,1},...,y_{i,n}=-1}}^1Q(x_{i,1},...,x_{i,l_i},\tilde{x}_{i,1},...,\tilde{x}_{i,n-l_i})y_{i,1}...y_{i,n}\left(\sum_{\substack{s_0,...,s_{l_i}=0\\ t_2,...,t_n=0}}^1G_{s_0...s_{l_i}t_2...t_n} y_{i,1}^{s_0}\left(\prod_{v=1}^{l_i}x_{i,v}^{s_v}\right)\left(\prod_{w=2}^n m_{i,w}^{t_w}\right)\right)
\\
\nonumber
& = & \frac{1}{2^{n-1}\Gamma}\sum_{\substack{x_{i,1},...,x_{i,l_i}=-1\\ \tilde{x}_{i,1},...,\tilde{x}_{i,n-l_i}=-1}}^1Q(x_{i,1},...,x_{i,l_i},\tilde{x}_{i,1},...,\tilde{x}_{i,n-l_i})\left(\sum_{s_1,...,s_{l_i}=0}^1G_{1s_1...s_{l_i}1...1}\left(\prod_{v=1}^{l_i}x_{i,v}^{s_v}\right)\right)\prod_{w=2}^n\left(\sum_{y_{i,w}=-1}^1y_{i,w} m_{i,w}\right)
\\
\nonumber
& = & \frac{1}{\Gamma}\sum_{\substack{x_{i,1},...,x_{i,l_i}=-1\\ \tilde{x}_{i,1},...,\tilde{x}_{i,n-l_i}=-1}}^1Q(x_{i,1},...,x_{i,l_i},\tilde{x}_{i,1},...,\tilde{x}_{i,n-l_i})\left(\sum_{s_1,...,s_{l_i}=0}^1G_{1s_1...s_{l_i}1...1}\left(\prod_{v=1}^{l_i}x_{i,v}^{s_v}\right)\right)\\ \nonumber
& & \prod_{w=2}^n\left(\sum_{r_{1},...,r_{l_w}=0}^1B^{(i,w)}_{1r_{1}...r_{l_w}}\prod_{u=1}^{l_w}x_{w,u}^{r_{u}}\right)
\end{eqnarray}

Now notice that from equation \ref{Message N parts} we have that:
\begin{eqnarray}
\sum_{y_{i,w}=-1}^1y_{i,w}m_{i,w} = 2\sum_{r_{1},...,r_{l_w}=0}^1B^{(i,w)}_{1r_{1}...r_{l_w}}\prod_{u=1}^{l_w}x_{w,u}^{r_{u}}.
\end{eqnarray}
Then, given that $m_{i,w}\in\{-1,1\}$, it must hold that:
\begin{eqnarray}
\label{Modulo inequality message}
\left|\sum_{r_{1},...,r_{l_w}=0}^1B^{(i,w)}_{1r_{1}...r_{l_w}}\prod_{u=1}^{l_w}x_{w,u}^{r_{u}}\right|\leq 1.
\end{eqnarray}

The same argument can be applied on equation \ref{Guess N parts} to get the relation:
\begin{eqnarray}
\label{Modulo Guess}
\left|\sum_{s_1,...,s_{l_i}=0}^1G_{1s_1...s_{l_i}1...1}\left(\prod_{v=1}^{l_i}x_{i,v}^{s_v}\right)\right|\leq 1
\end{eqnarray}

These two relation allow us to define the correlation functions given by:
\begin{eqnarray}
E_{x_{i,1}...x_{i,l_i}} = \sum_{s_1,...,s_{l_i}=0}^1G_{1s_1...s_{l_i}1...1}\left(\prod_{v=1}^{l_i}x_{i,v}^{s_v}\right),
\end{eqnarray}
and
\begin{eqnarray}
E_{x_{w,1}...x_{w,l_w}} = \sum_{r_{1},...,r_{l_w}=0}^1B^{(i,w)}_{1r_{1}...r_{l_w}}\prod_{u=1}^{l_w}x_{w,u}^{r_{u}}
\end{eqnarray}

Thus the fidelity reads:
\begin{eqnarray}
(f,G_i) & = & \frac{1}{\Gamma}\sum_{\substack{x_{i,1},...,x_{i,l_i}=-1\\ \tilde{x}_{i,1},...,\tilde{x}_{i,n-l_i}=-1}}^1Q(x_{i,1},...,x_{i,l_i},\tilde{x}_{i,1},...,\tilde{x}_{i,n-l_i})E_{x_{i,1}...x_{i,l_i}}\prod_{w=2}^nE_{x_{w,1}...x_{w,l_w}}
\\ \nonumber
& \leq & \frac{B_n^C}{\Gamma}
\end{eqnarray}

This implies that the success probability of at most:
\begin{eqnarray}
\mathcal{P}_{i} = \frac{1}{2} + \frac{B_n^c}{\Gamma}, \;\;\;\forall i;
\end{eqnarray}
being always less or equal to the obtained with the presented strategy, proving it to be optimal.


%

\end{document}